\documentclass[epj]{svjour}

\usepackage{graphicx}% Include figure files
\usepackage{dcolumn}% Align table columns on decimal point
\usepackage{bm}% bold math

\begin{document}

\title{Modelling disorder: the cases of wetting and DNA denaturation}

\author{S. Ares\inst{1}\and A. S\'anchez\inst{2}}
\institute{
Max Planck Institut f\"ur Physik komplexer Systeme,
N\"othnitzer Str. 38, 01187 Dresden, Germany. \email{saul@mpipks-dresden.mpg.de}
\and Grupo Interdisciplinar de Sistemas Complejos
(GISC) and Departamento de Matem\'aticas, Universidad Carlos III de
Madrid, Avenida de la Universidad 30, 28911 Legan\'es, Madrid,
Spain;\\
Instituto de Biocomputaci\'on y F\'\i sica de Sistemas
Complejos, Universidad de Zaragoza, 50009 Zaragoza, Spain
}%


\abstract{We study the effect of the composition of the genetic sequence on
the melting temperature of double stranded DNA, using some simple
analytically solvable models proposed in the framework of the
wetting problem. We review previous work on disordered versions of
these models and solve them when there were not preexistent
solutions.
We check the solutions with Monte Carlo simulations and
transfer matrix numerical calculations.
We present numerical evidence that suggests
that the logarithmic corrections to the
critical temperature due to disorder, previously found in RSOS models,
apply more generally to ASOS and continuous models.
The agreement
between the theoretical models and experimental data shows that,
in this context,
disorder should be the crucial ingredient of any
model while other aspects may
be kept very simple, an approach that can be useful
for a wider class of problems.
Our work has also implications for the existence of correlations in
DNA sequences.
\PACS{
{87.15.-v}{Biomolecules: structure and physical properties}\and
{68.35.Rh}{Phase transitions and critical phenomena}\and
{05.40.-a}{Fluctuation phenomena, random processes, noise, and Brownian motion}
}}

\maketitle

\section{Introduction}
\label{sec:int}

The study of disorder in physical systems has proved fruitful not
only for better understanding of complex systems, but also because
new and powerful methods have been developed in order to cope with
the problems posed by disordered systems. Thus, tools developed in
the context of physical problems have been applied to the study of
biological systems, where disorder (or, in general, inhomogeneity)
is ubiquitous. In fact, inhomogeneity is frequently the most
important part in a given biological function. This is the case,
for instance, with proteins, where the biological function is
given by the way they fold, which in turn depends on the sequence
of aminoacids that compose the protein.

The example we are interested in is also of fundamental biological
importance: the genetic sequence contained in the DNA molecule.
Theoretical studies of the thermal denaturation (or melting)
transition of DNA have resorted to different kinds of models:
Ising-like models such as those introduced by Poland and Scheraga
\cite{PS}, or Hamiltonian models including physical interactions,
e.g. the model proposed by Peyrard and Bishop \cite{PB,dpb} and
other nonlinear models \cite{yaku}. Recently, the inclusion of the
genetic sequence in this model has shown the physical importance
of functional sites for transcription
\cite{NAR,preprintmichel} or the theoretical explanation
\cite{miprl} of the experimentally observed \cite{Zocchi}
denaturation bubbles and cooperative effects in the melting
process. In the spirit of the original Peyrard-Bishop model
\cite{PB}, we will see that two simple models proposed in 1981 by
Chui and Weeks \cite{chui}
(similar work was carried at the same time by van Leeuwen and Hilhorst
\cite{leeuwen})
and Burkhardt \cite{burkhardt} for the
study of the wetting transition are able, in spite of their
simplicity, to characterize the effect of the disorder introduced
by the genetic sequence on the melting temperature. In the
remainder of the paper, we discuss these models and apply them to
the DNA melting problem. From the specific
viewpoint of the models considered, our numerical results
support the idea that
in the ASOS model and in the continuous model the corrections to the critical
temperature introduced by disorder are logarithmic as in the RSOS model.
From a more general viewpoint, as we will see, in
spite of the extreme simplicity of the models, the results are in good
semi-quantitative
agreement with the experiments. This suggests that
in this system the most relevant
ingredient is disorder, allowing for modeling other
aspects of the problem in a crude
but efficient way while still obtaining a
generally correct description. We discuss this
point in more detail in the last section and
suggest that it may be useful for models
in other contexts. Important conclusions
regarding the nature of possible correlations
in DNA are also drawn.

\section{Homogeneous Chui-Weeks and Burkhardt models}
\label{sec:chui}

In 1981, Chui and Weeks \cite{chui} (see also Ref. \cite{leeuwen})
proposed a wetting model to study the depinning of an
interface from an attractive substrate. The version of their model in which we
are interested can be written in terms of a Hamiltonian defined in a one
dimensional lattice with periodic boundary conditions:
\begin{equation}
\label{eq:4:hamil_cw}
\mathcal{H}=\sum_{i=1}^N\Big\{J|n_{i+1}-n_i|-B\delta_{n_i,0}\Big\},
\end{equation}
where $n_i$ are integer variables with positive or zero values, and
$N$ is the total number of nodes in the lattice. The variables
$n_i$ are the distance between the substrate and the interface
position at site $i$. The first term in the Hamiltonian can be
understood as a surface tension, where $J$ is the coupling
constant between nearest neighbors. The second term is a potential
which binds the interface to the substrate at $n_i=0$, and $B$ is
the strength of this attraction. Chui and Weeks showed
analytically that this model presents a thermodynamic phase
transition (see \cite{1d} for a recent review on the existence of
one-dimensional phase transitions) between a bound interface at
low temperatures and a free one at high temperatures.The critical
condition for this transition is given by:
\begin{equation}
\label{eq:cw_critical}
e^{-\beta_c B}=1-e^{-\beta_c J},
\end{equation}
where $\beta_c=1/k_BT_c$.

In the same year as Chui and Weeks, Burkhardt \cite{burkhardt}
proposed a model that is the continuous version of the Chui-Weeks
discrete model. Its Hamiltonian is:
\begin{equation}
\label{eq:4:hamil_burk}
\mathcal{H}=\sum_{i=1}^N\Big\{J|y_{i+1}-y_i|+V(y_i)\Big\},
\end{equation}
where now $0\leq y_i<\infty$ are real variables and the potential
$V(y_i)$ has the form:
\begin{eqnarray}
\label{eq:pot_burk}
V(y_i)=
\left \{ \begin{array}{ll}
-B, & y_i \leq R,\\
0, & y_i>R.
\end{array} \right.
\end{eqnarray}
The model is equivalent to the one proposed by Chui and Weeks, but
now the variables are continuous and the substrate attraction has a
finite range given by the constant $R$.

Burkhardt showed analytically that this model has the same kind of
phase transition than the discrete model. The critical condition now
is:
\begin{equation}
\label{eq:4:burk_critical}
\beta_c J R=(e^{\beta_c B}-1)^{-1/2}
\tan^{-1}\left[(e^{\beta_c B}-1)^{-1/2}\right].
\end{equation}

\section{Disorder}

Forgacs {\em et al} \cite{forgacs} (see also \cite{derrida})
extended the discrete model in its restricted
(RSOS) version, where the
restriction $|n_{i+1}-n_i|\leq 1$ is enforced, to include
disorder in the potential. {The disorder is introduced by letting
the potential parameter $B$ in Eq. (\ref{eq:4:hamil_cw}) be site
dependent, uncorrelated and given by a distribution $P(b)$.
Depending on whether the partition function is averaged over the
disorder (in which case the free energy is
$F_a=-k_BT\log\overline{\mathcal{Z}}$) or the average is done
directly over the free energy
($F_q=-k_BT\overline{\log\mathcal{Z}}$) the disorder is said to be
{\em annealed} or {\em quenched}. In the case of annealed
disorder} the factor $\exp(\beta B)$ which appears at the
beginning in the calculation of the solution of the model should
be substituted by the average \cite{forgacs,derrida}:
\begin{equation}
\label{eq:cw_average}
e^{\widetilde{\beta B}}=
\int_{-\infty}^{\infty}\textrm{d}bP(b)e^{\beta b}.
\end{equation}
Instead of the restricted version of the model (RSOS) studied
in references \cite{forgacs,derrida}, we are interested in the
unrestricted version (ASOS) where the difference $|n_{i+1}-n_i|$
can take any positive (or zero) value. To our knowledge, the ASOS
model with disorder was not studied in previous references.
Following a procedure similar to that for the RSOS case, the new
critical condition for the ASOS model with annealed disorder can
be found to be:
\begin{equation}
\label{eq:cw_critical_des}
\left(e^{\widetilde{\beta B}}\right)^{-1} =1-e^{-\beta_c J}.
\end{equation}
In the case of a dichotomous disorder which can take a value $B_1$
with probability $p$ and a value $B_2$ with probability $1-p$, this
condition reads:
\begin{equation}
\label{eq:cw_critical_des2} \frac{1}{pe^{\beta B_1}+(1-p)e^{\beta
B_2}} =1-e^{-\beta_c J}.
\end{equation}
Nevertheless, the kind of disorder relevant in a DNA molecule is
quenched disorder, as the genetic sequence does not change with
time. Although in reference \cite{forgacs} the conclusion is
reached that the annealed and the quenched cases yield the same
critical temperature, later work \cite{derrida} (reinforced by work on a related
model defined on a one-dimensional continuum \cite{bhat})
showed that there
is a logarithmic difference between the critical temperature for
both types of disorder, which vanishes when the disorder goes to
zero. For the values of the parameters we are interested in (see
section \ref{sec:dna}), the difference between the critical
temperature predicted for both types of disorder is several orders
of magnitude smaller than the temperatures themselves, and hence
not measurable by usual means.

Although references
\cite{forgacs,derrida} work for the restricted version of the
model, we believe that the same phenomenology appears in the
unrestricted one we are using. Subsequently, we will use Eq.\
(\ref{eq:cw_critical_des2}) as an approximation for quenched
disorder, and we will check afterwards the validity of this
approximation by two different numerical approaches: Monte Carlo
simulation of the model and numerical evaluation of the transfer
matrix.

As for the Burkhardt model, to our knowledge no attempt has been
made to study it with uncorrelated disorder in the potential in
the same fashion we have seen for the discrete model. Burkhardt
\cite{burkhardt2} made a study using a corrugated potential, but
constructed in a deterministic and periodic way. {In reference
\cite{nowa} both the discrete and the continuous models are
studied with a somewhat more general kind of disorder in the
potential, that also has to be periodic.} However, uncorrelated
disorder can be studied in the continuous model in the same way
that in the discrete one. If the constant $B$ is site dependent
and given in each site by a probability distribution $P(b)$, {for
annealed disorder} the factor $\exp(\beta B)$ in Burkhardt's
calculation \cite{burkhardt} should be substituted by:
\begin{equation}
\label{eq:4:Ubar}
e^{\widetilde{\beta B}}=
\int_{-\infty}^{\infty}\textrm{d}b\
P(b)e^{\beta b}.
\end{equation}
This leads to the following critical condition in the presence of
{annealed} disorder:
\begin{equation}
\label{eq:4:burk_critical_des}
\beta_c J R=(e^{\widetilde{\beta_c B}}-1)^{-1/2}
\tan^{-1}\left[(e^{\widetilde{\beta_c B}}-1)^{-1/2}\right].
\end{equation}
In the case of a dichotomous disorder as we saw in the discrete
model, the factor $\exp(\widetilde{\beta B})$ is:
\begin{equation}
\label{eq:burk_critical_des2}
e^{\widetilde{\beta B}}=pe^{\beta B_1}+(1-p)e^{\beta B_2}.
\end{equation}
{Again, as in the discrete model, we will assume this formula to
be a good approximation also for quenched disorder, and check
later this hypothesis by Monte Carlo simulations and numerical
evaluation of the transfer operator of the model. }

\section{The Peyrard-Bishop model of DNA}
\label{sec:PB} One of the most successful models in biophysics is
the Peyrard-Bishop model, which describes
in a simplified way (including only one relevant degree of freedom
for each base pair, namely the distance between the two bases)
the dynamics and
statistical mechanics of the DNA molecule. The original
formulation by Peyrard and Bishop can be written in terms of the
following Hamiltonian \cite{dpb2}, disregarding the kinetic term
in which we are not interested in this work:
\begin{equation}
\label{eq:dpb_hamil}
\mathcal{H}=\sum_{i=1}^N\Big\{\frac{J}{2}(y_{i+1}-y_i)^2+V(y_i)\Big\},
\end{equation}
where $V(y_i)=B(e^{-Ry_i}-1)^2$ is a Morse potential, $B$
giving the strength of the potential and $R$ the width of the attracting well of
the potential. The variables $y_i$ can take any real value, and they represent
the difference of the actual
distance between the two bases in base pair $i$ and their
equilibrium distance. The harmonic coupling represents the rigidity of the
molecule, due to in part to the stacking interaction between consecutive base
pairs. The Morse potential represents the hydrogen bonds between the two bases
in a base pair.

\begin{figure}
\vspace*{2mm}
\includegraphics[width=8.5cm]{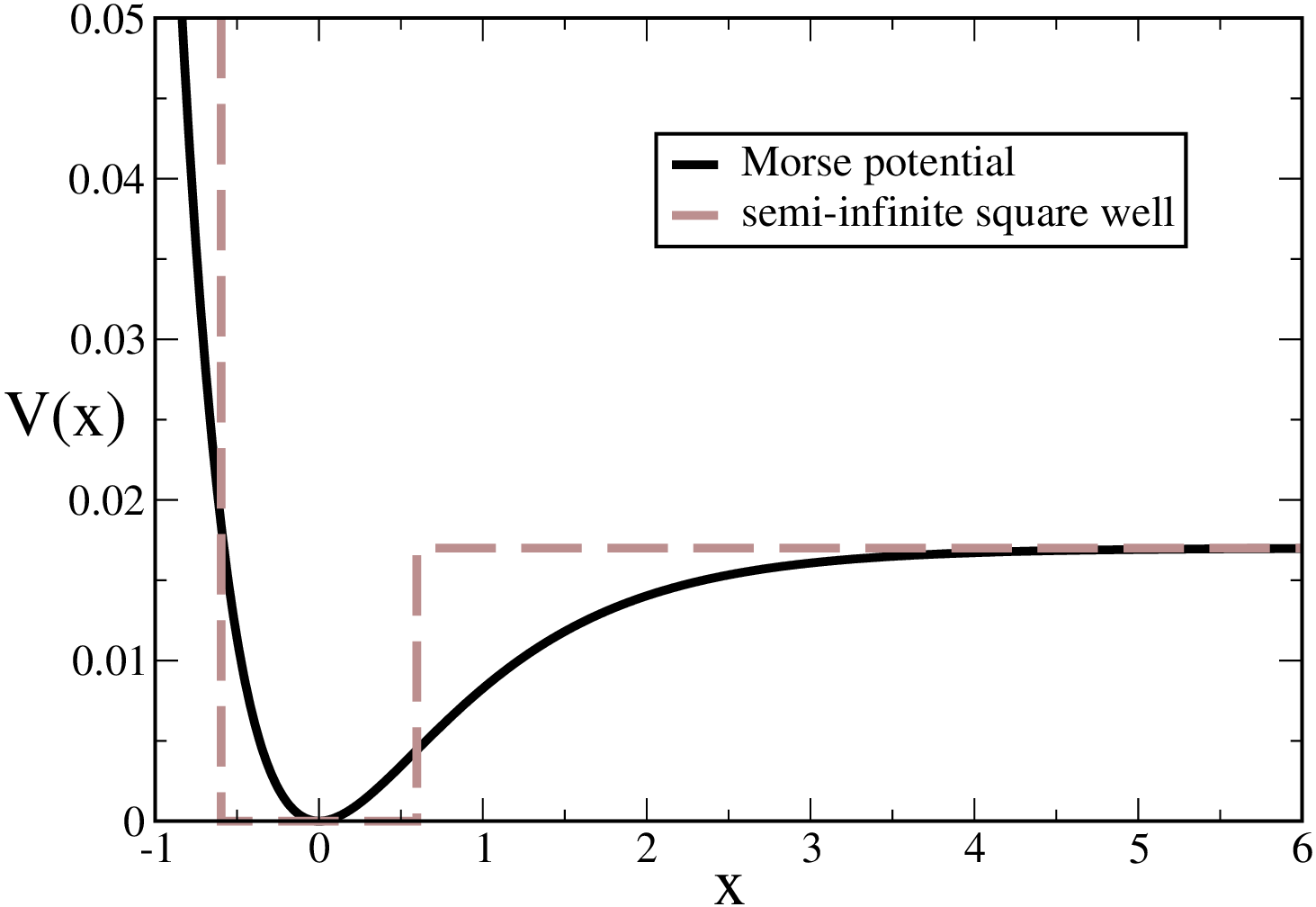}
\caption{\label{fig:morse}Comparison between a Morse potential with parameters
$B=0.017$ and $R=1.195$ with a semi-infinite square well with the same
parameters. The latter has been displaced in order to help the eye.}
\end{figure}
Note that the Peyrard-Bishop model is formally very similar to the
Burkhardt model. The difference in the type of coupling between
both of them (absolute value in Burkhardt, harmonic in
Peyrard-Bishop) does not introduce qualitative differences. The
Morse potential and the semi-infinite square well in the Burkhardt
model are qualitatively very similar, as can be seen in Figure
\ref{fig:morse}. These facts make us expect the Burkhardt model to
display the same qualitative behavior than Peyrard-Bishop's. This
means that we can use the simple Burkhardt model, and the even
simpler Chui-Weeks model, its discrete counterpart, to study some
aspects of the DNA molecule. In this way we can take advantage
from the fact that from these models analytical information
can be extracted even
when disorder is included.

The dichotomous disorder we introduced above is just the kind of
effect that the genetic sequence introduces in the physics of the
DNA molecule: the existence of only two kinds of base pairs
(adenine-thymine, A-T, and guanine-cytosine, G-C) with different
number of hydrogen bonds (2 for A-T and 3 for G-C) implies that we
should take into account two different values for the potential
strength $B_i$, depending on whether the site $i$ corresponds to
an A-T pair or to a G-C pair. For simplicity, in this work we make
only the strength of the potential site dependent, keeping its
width $R$ constant on all sites.

\section{DNA melting}
\label{sec:dna} 
\begin{figure}
\vspace*{2mm}
\includegraphics[width=8.5cm]{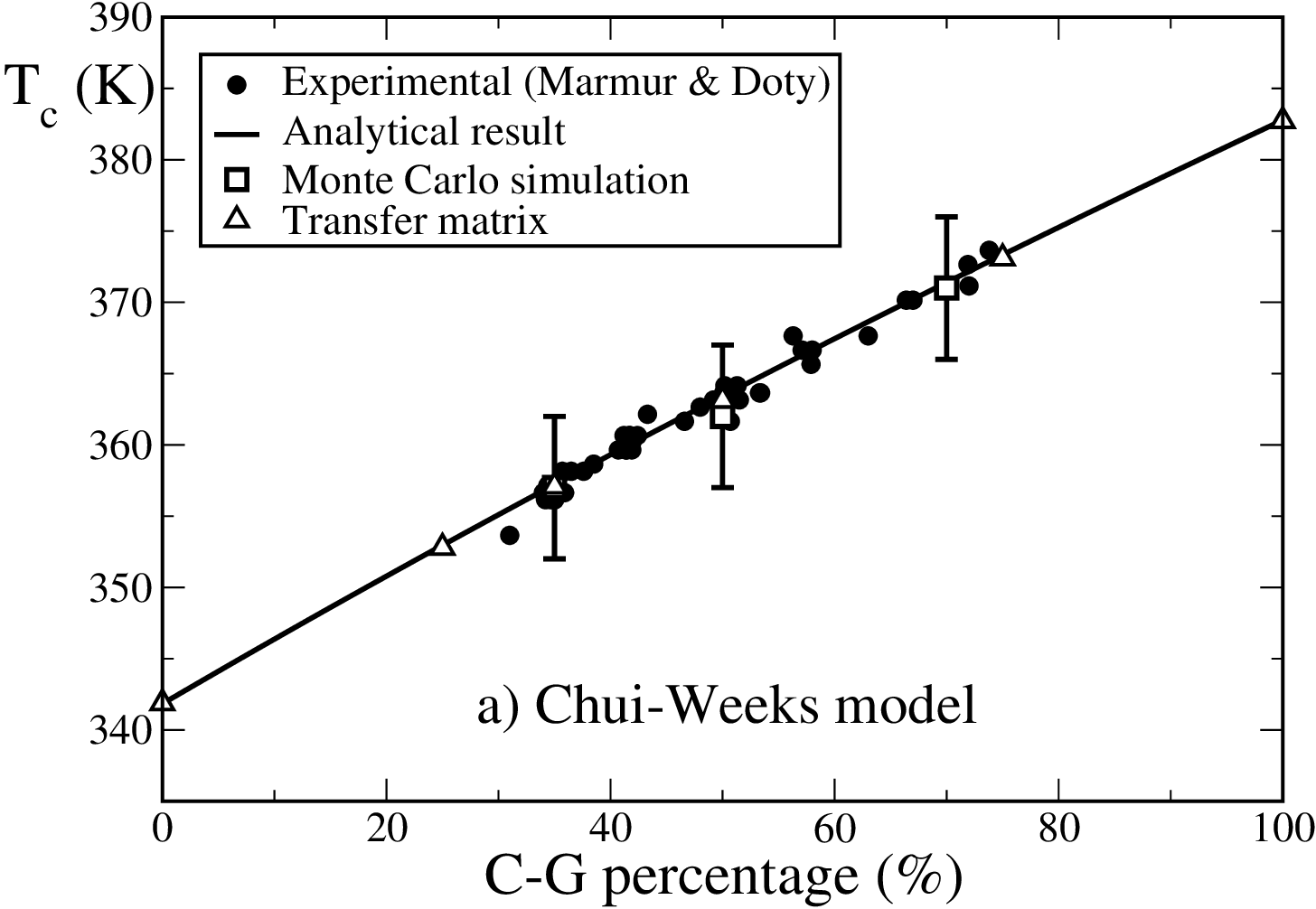}
\includegraphics[width=8.5cm]{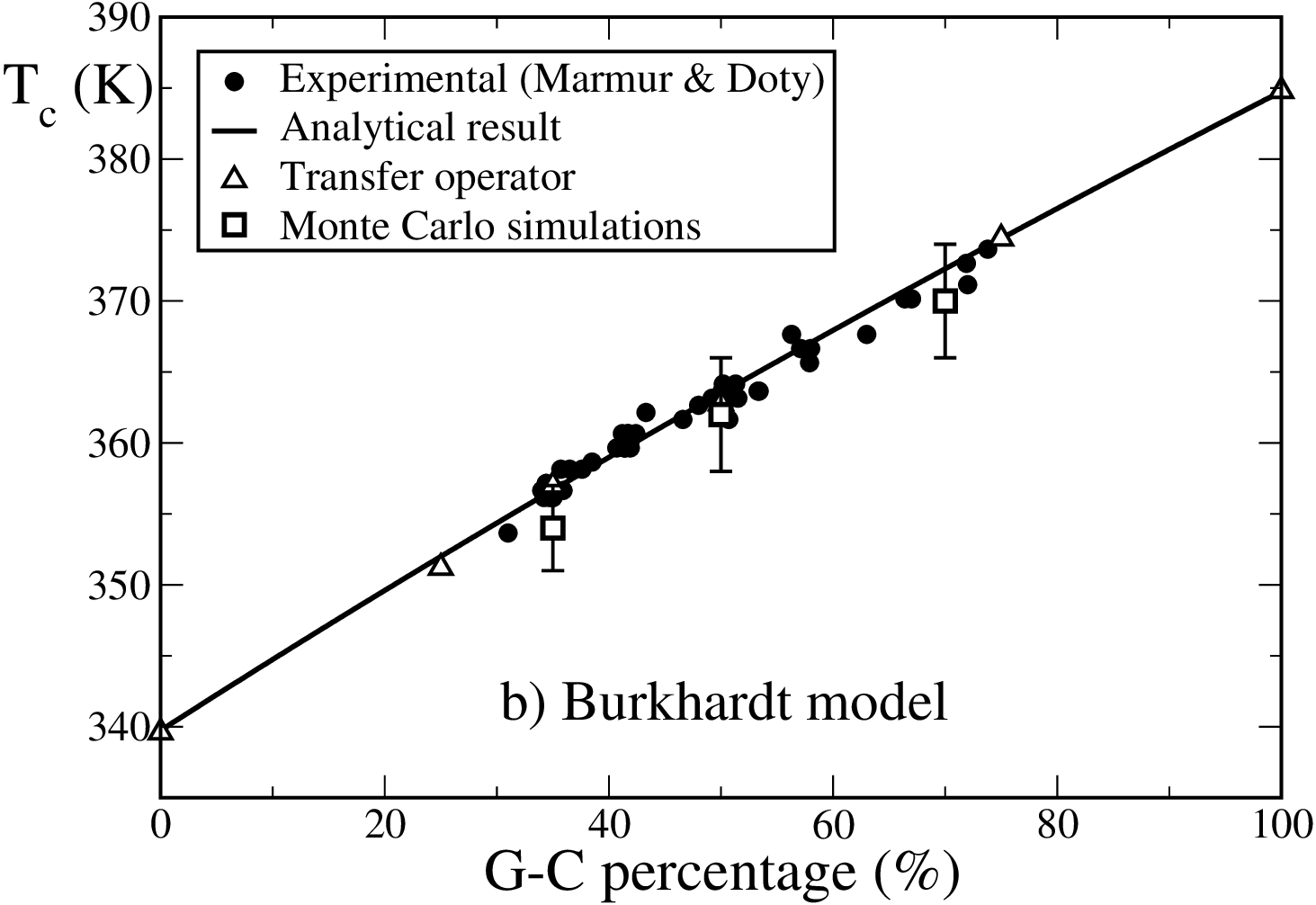}
\caption{\label{fig:ADN} Comparison of the analytical results for the annealed
versions of the models
(Eq. (\ref{eq:cw_critical_des2}) in figure a and Eq.
(\ref{eq:burk_critical_des2})
in figure b)
and the numerical results (Monte Carlo and transfer
operator) for the quenched versions. Comparison is made as well with
experimental results from Ref. \cite{marmur} for
the melting temperature as a function of G-C concentration.
Note that although they look linear, the predictions of Eqs.
(\ref{eq:cw_critical_des2}) and (\ref{eq:burk_critical_des2}) are not linear
fits.}
\end{figure}

The thermal denaturation of DNA is a phase
transition where the effect of the genetic sequence can not be
ignored \cite{miprl}. A first (rough)
approximation, leaving out the
structure in the sequence, is making the hypothesis that a long
enough DNA molecule can be treated as an uncorrelated sequence of
A-T and G-C pairs, with a fixed concentration of pairs of each
type. Then, using the Burkhardt model as a simplification of the
Peyrard-Bishop model, we can obtain the critical temperature for a
given concentration from Eq. (\ref{eq:4:burk_critical_des}). An
even stronger simplification is to use the Chui-Weeks model, in
which case it is Eq. (\ref{eq:cw_critical_des2}) the one that
gives the critical temperature.
This makes the DNA molecule a useful playground to test
experimentally the predictions of including disorder in the wetting
models studied.
In Figure \ref{fig:ADN}, we show
the comparison between experimental data in Ref. \cite{marmur} and
the prediction of Eqs. (\ref{eq:cw_critical_des2}) and
(\ref{eq:4:burk_critical_des}), Monte Carlo simulations \cite{MC} of the
models using the Metropolis algorithm \cite{Metropolis} and
parallel tempering \cite{newman,iba}, and the numerical evaluation
of the transfer operator \cite{BsG} of the models. The numerical
studies confirm that Eqs. (\ref{eq:cw_critical_des2}) and
(\ref{eq:4:burk_critical_des}) are good approximations for
quenched disorder, as we can see by the coincidence of the
analytical results for annealed disorder and numerical
ones for quenched disorder.
This is not strange, since Eq. (1.7) in reference \cite{derrida}
predicts that, for the discrete RSOS model, the difference between
$e^{\widetilde{\beta B_c}}$ of the random model and
$e^{\beta B_c}$ of the pure one is of the order of $10^{-24}$ for the
parameters we use,
and similar behavior is expected in the ASOS and continuous models
we study.
The parameters used for both
models are: $J=0.03$ eV, $B_{G-C}=0.017$ eV, $B_{A-T}=0.0132$ eV
and $R=1.195$ \AA. These values are of the order of magnitude of
the parameters accepted for the Peyrard-Bishop model \cite{campa},
and, as can be seen in the figure,
using them the wetting models studied
reproduce correctly the experimental dependence of $T_c$ on the
sequence composition.

\section{Conclusions}
\label{sec:univ}

%For the RSOS version of the Chui and Weeks model, Ref. \cite{derrida} showed
%that the critical temperature of the model with quenched disorder has
%logarithmic corrections with respect to the critical temperature of the model
%with annealed disorder.
%We have shown that for the ASOS version of the model analytical calculations
%of the critical temperature with annealed disorder show good agreement with Monte Carlo simulations and
%transfer integral calculations using quenched disorder, hence indicating that
%the possible difference is small, compatible with a logarithmic correction.
%Moreover, we have calculated analytically the critical temperature with
%annealed disorder for a continuous version of the model due to Burkhardt. We have
%also realized numerical calculations for this model with quenched disorder,
%showing again that the difference between the critical temperature of both
%models is small and compatible with logarithmic corrections.

In this work we have revisited two models of wetting, the first
introduced by Chui and Weeks and the second by Burkhardt. We have
presented the treatment made for the ASOS Chui-Weeks model with
disorder and developed an original treatment for the Burkhardt
model, finding the critical conditions for both models when the
disorder is annealed. Two independent numerical approaches, namely Monte
Carlo simulation and evaluation of the transfer operator of the
models, confirm that the expressions obtained for annealed disorder are
good approximations for the critical conditions with quenched
disorder. This suggests that, as in the discrete RSOS model, in the
ASOS model and in the continuous model the corrections to the critical
temperature introduced by disorder are also logarithmic.
Using this two models as simplified versions of the
Peyrard-Bishop model of DNA, we have been able to reproduce the
dependence on the genetic sequence of the DNA melting temperature.

The agreement between the results of this simple models
(one of them is even of discrete nature) and experimental results
suggests that,
for the class of inhomogeneous models studied here,
the
effect of the disorder
can be characterized using simplified interactions, as long as the
disorder itself is properly taken into account.
This indicates that a
theoretical model where the disorder is introduced in a
proper and realistic way can reproduce correctly the effect of
disorder, even if the other interactions in the model are not
included in a detailed way. 
Indeed, we
have seen that very simple and unrealistic models are enough to
display the dependence of the melting temperature of DNA with the
concentration of each kind of base pairs. These models do not even
have a phase transition of the correct order: It is continuous,
while experimentally the melting transition is seen to be first
order. A further refinement of the Peyrard-Bishop model, which is
more realistic but at the same time more complex, as it includes
an anharmonic coupling \cite{dpb}, shows a phase transition of the
correct order. However, the description of the effects of the
sequence in the general level we have used in this work does not
improve much by taking this more realistic model. We have checked,
using Monte Carlo Metropolis simulations as those described in
\cite{miprl}, that the correct relation between melting
temperature and G-C concentration is also recovered by this model
\cite{unpub}, which stands as yet another successful experimental
test of the Peyrard-Bishop model with the anharmonic coupling, and
strengthens our claim that the quite simple Chui-Weeks or Burkhardt
models where enough to display this phenomenology.
Further
research about the comparison of other properties as computed from
the simple models with
the experiments would be necessary to ascertain the degree of
usefulness of the models.

\begin{figure}
\vspace*{2mm}
\includegraphics[width=8.5cm]{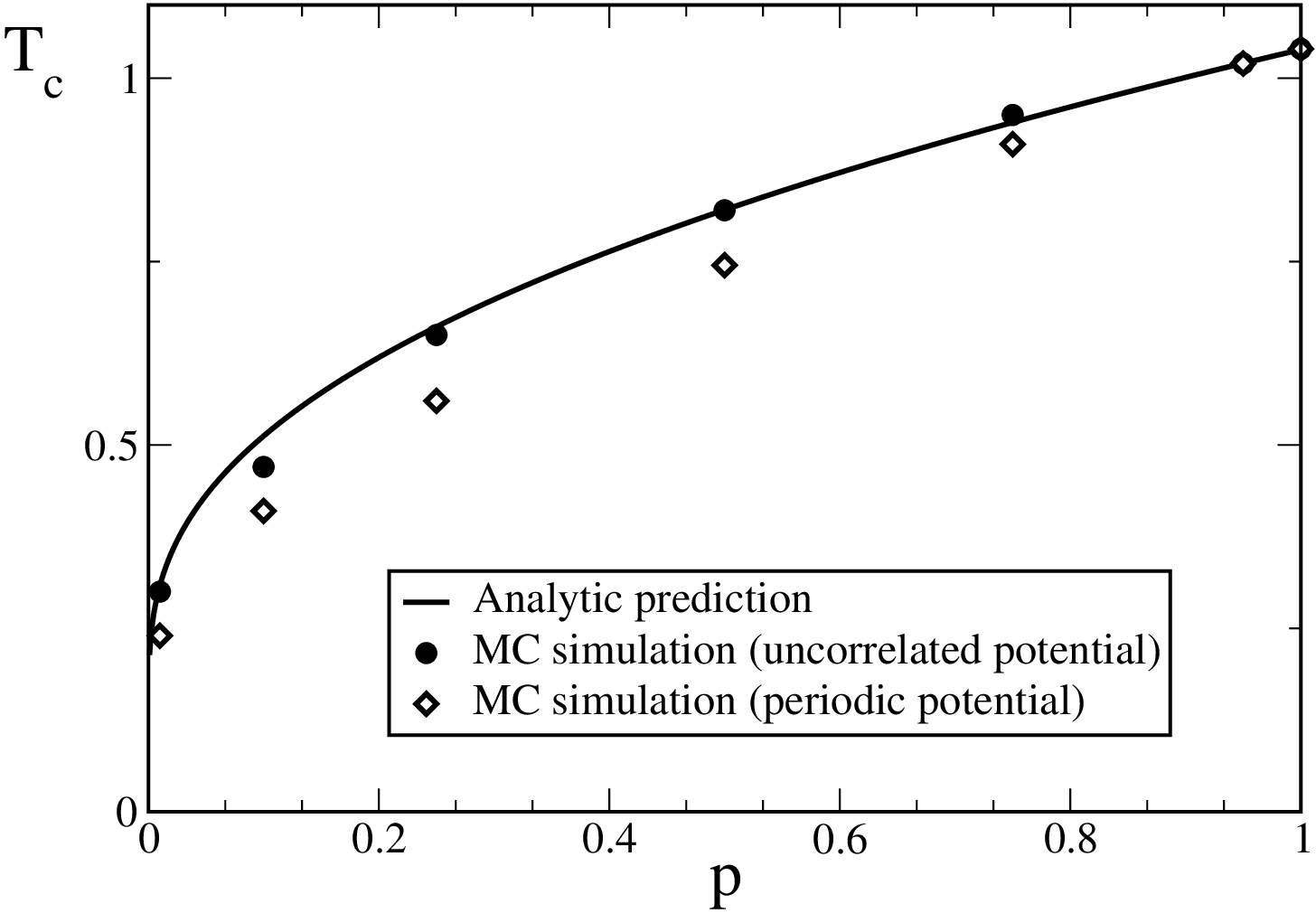}
\caption{\label{fig:per} Critical temperature of the Chui-Weeks
model with dichotomous disorder as a function of the fraction of
sites $p$ with potential strength $B_1$. Parameters: $B_1=1$,
$B_2=0$, $J=0.5$.}
\end{figure}
This approach, based on choosing disorder as the factor to be modelled more
accurately,
has proved useful in other contexts other than the one explored here,
such as, e.g., protein folding, where a huge huge amount of work has been
done using oversimplified models that, nevertheless,
have given good results, that even resist quantitative comparison
with experiments \cite{xxx}. Another example can be found in
problems involving complex networks, where real problems can be
reproduced with simplified interactions or rules granted that the
underlying network structure is correctly described \cite{yyy}.
Finally, spin glasses models show how a simple description with
complex disorder gives powerful insight of natural phenomena
\cite{binder}. We believe that other problems in the field of complex systems
may benefit from these ideas as, generally speaking, designing
simple models as those considered here with a correct description
of the disorder may be a very efficient way to obtain rapidly
approximate results that may guide further, more detailed studies.

%The Peyrard-Bishop model as presented in Eq. (\ref{eq:dpb_hamil})
%is already a rough simplification: to have an improved model an
%anharmonic coupling is needed \cite{dpb}. The model given by Eq.
%(\ref{eq:dpb_hamil}) does not even have a phase transition of the
%correct order: while experimentally and in the version of the
%model with anharmonic coupling the melting transition is first
%order like, the model in Eq. (\ref{eq:dpb_hamil}) presents a
%continuous phase transition, as also the Chui-Weeks and Burkhardt
%models do. This is a strong reinforcement of our disorder
%universality hypothesis, because it shows that the example we have
%chosen to explain the concept is in fact a very poor model far
%from being realistic, and nevertheless it has been able to display
%correctly the effects of disorder. Of course, the more realistic
%model proposed in reference \cite{dpb} also reproduces correctly
%the dependence of the melting temperature on the concentration of
%G-C \cite{unpub}, what constitutes in fact yet another successful
%experimental test of the Dauxois-Peyrard-Bishop model.

Finally, another conclusion arises from the fact that
all the DNA sequences used in reference \cite{marmur} are natural
sequences coming from living organisms. The fact that assuming
uncorrelated sequences we have been able to compute the melting
temperature means that, in fist approximation, it is reasonable to
consider natural that DNA, at least long enough sequences,
melts as if it were effectively
uncorrelated.
In Figure \ref{fig:per} we see a comparison of
simulation results for uncorrelated sequences and periodic
sequences. We see that the analytic expression works well for the
uncorrelated case, but there is a degree of deviation for the
periodic one. Therefore, we do not expect our approach to hold for
quasi-periodic or highly correlated sequences.
Interestingly, this implies that if there actually exist correlations in the
sequences used in the experiments in \cite{marmur},
they do not affect the melting temperature, which remains unchanged with
respect to the melting temperature of purely random sequences.
We want
to remark again that we have found the same results
using the Peyrard-Bishop model
with anharmonic coupling \cite{dpb}. This strengthens the claim that
including correlations in the sequence is not crucial for obtaining the
correct critical temperature.
This observation does not challenge the existence of correlations,
even long ranged ones \cite{zzz},
in DNA: it just states that they are weak enough
to not have an appreciable effect over the melting temperature of
the molecule, although they may be of great biological significance.

\begin{acknowledgement}
This work has been supported by the Ministerio de Ciencia y
Tecnolog\'\i a of Spain through grants MOSAICO
and NAN2004-09087-C03-03
and Ingenio Mathematica (i-MATH) (AS)
and by Comunidad de Madrid grant SIMUMAT-CM (AS).
\end{acknowledgement}


\begin{thebibliography}{30}

\bibitem{PS} D. Poland and H.A. Scheraga, J. Chem. Phys. {\bf 45}, 1456 (1966);
{\bf 45}, 1464 (1966); D. Poland, Biopolymers {\bf 73}, 216 (2004);
C. Richard and  A. J. Guttmann, J. Stat. Phys. {\bf 115}, 925 (2004)
\bibitem{PB} M.\ Peyrard and A.\ R.\ Bishop, Phys.\ Rev.\ Lett.\ {\bf 62}, 2755
(1989)
\bibitem{dpb} T.\ Dauxois, M.\ Peyrard and A.\ R.\ Bishop, Phys.\ Rev.\ E
{\bf 47}, R44 (1993);
T.\ Dauxois and M.\ Peyrard, Phys.\ Rev.\ E {\bf 51}, 4027 (1995)
\bibitem{yaku} L.\ V.\ Yakushevich, {\em Nonlinear Models of DNA}
(2nd edition, Wiley, 2004)
\bibitem{NAR} C. H. Choi, G. Kalosakas, K. {\O}. Rasmussen, M. Hiromura,
A. R. Bishop and A. Usheva,
Nucleic Acids Res. {\bf 32}, 1584 (2004);
G. Kalosakas, K. {\O}. Rasmussen, A. R. Bishop, C.H. Choi, and A. Usheva,
Europhys. Lett. {\bf 68}, 127 (2004)
\bibitem{preprintmichel}  T.\ S.\ van Erp, S.\ Cuesta-L\'opez,
J.-G.\ Hagmann and M.\ Peyrard,
Phys.\ Rev.\ Lett.\
{\bf 95}, 218104 (2005)
\bibitem{miprl} S. Ares, N. K. Voulgarakis, K. {\O}. Rasmussen and A. R. Bishop,
Phys.\ Rev.\ Lett.\ {\bf 94}, 035504 (2005)
\bibitem{Zocchi} A.\ Montrichok, G.\ Gruner and G.\ Zocchi, Europhys.\ Lett.\
{\bf 62}, 452 (2003);
Y.\ Zeng, A.\ Montrichok and G.\ Zocchi, Phys.\ Rev.\ Lett.\
{\bf 91}, 148101 (2003);
Y.\ Zeng, A.\ Montrichok and G.\ Zocchi, J.\ Mol.\ Biol.\
{\bf 339}, 67 (2004)
\bibitem{chui} S.\ T.\ Chui and J.\ D.\ Weeks, Phys.\ Rev.\ B {\bf 23},
R2438 (1981)
\bibitem{leeuwen} J.\ M.\ J.\ van Leeuwen and H.\ J.\ Hilhorst,
Physica A {\bf 107}, 319 (1981).
\bibitem{burkhardt} T.\ W.\ Burkhardt, J.\ Phys.\ A {\bf 14}, L63 (1981)
\bibitem{1d} J.\ A.\ Cuesta and A.\ S\'anchez, J.\ Stat.\ Phys.\
{\bf 115}, 869 (2004)
\bibitem{forgacs} G.\ Forgacs, J.\ M.\ Luck, Th.\ M.\ Nieuwenhuizen and H.\ Orland,
Phys.\ Rev.\ Lett.\ {\bf 57}, 2184 (1986);
J.\ Stat.\ Phys.\ {\bf 51}, 29 (1988)
\bibitem{derrida} B.\ Derrida, V.\ Hakim and J.\ Vannimenus, J.\ Stat.\ Phys.\ {\bf 66},
1189 (1992)
\bibitem{bhat} S.\ M.\ Bhattacharjee and S.\ Mukherji,
Phys.\ Rev.\ Lett.\ {\bf 70}, 49 (1993); S.\ Mukherji and S.\ M.\ Bhattacharjee,
Phys.\ Rev.\ E {\bf 48}, 3483 (1993)
\bibitem{burkhardt2} T.\ W.\ Burkhardt, J.\ Phys.\ A {\bf 31}, L549 (1998)
\bibitem{nowa} P. Nowakowski and M. Napi\'orkowski, J. Phys. A {\bf 38},
5885 (2005)
\bibitem{dpb2} T.\ Dauxois, M.\ Peyrard and A.\ R.\ Bishop,
Phys.\ Rev.\ E {\bf 47}, 684 (1993);
\bibitem{marmur} J.\ Marmur and P.\ Doty, J. Mol. Biol. {\bf 5}, 109 (1962)
\bibitem{MC} Monte Carlo simulations have been done following the procedure
described in \cite{MC2}. $5\cdot 10^5$ tries of replica exchange are used, and
between them each replica is simulated a number of Monte Carlo
steps equal to the energy autocorrelation time at the replica's temperature,
determined during previous simulations used for equilibration.
\bibitem{MC2} S. Ares, J. A. Cuesta, A. S\'anchez and R. Toral,
Phys. Rev. E {\bf 67}, 046108, (2003)
\bibitem{Metropolis} N.\ Metropolis, A.\ W.\ Rosenbluth, M.\ N.\ Rosenbluth,
A.~H.\ Teller and E.\ Teller,
J.\ Chem.\ Phys.\ {\bf 21}, 1087 (1953)
\bibitem{newman} M.\ E.\ J.\ Newman and G.\ T.\ Barkema,
{\em Monte Carlo Methods in Statistical Physics} (Oxford University,
Oxford, 1999)
\bibitem{iba} Y.\ Iba, Int.\ J.\ Mod.\ Phys.\ C {\bf 12}, 623 (2001)
\bibitem{BsG} S. Ares and A. S\'anchez, Phys.\ Rev.\ E {\bf 70}, 061607 (2004)
\bibitem{campa} A.\ Campa and A.\ Giansanti, Phys.\ Rev.\ E {\bf 58}, 3585 (1998)
\bibitem{unpub} S. Ares and A. S\'anchez, {\em unpublished data}.
\bibitem{xxx} B. H. Park and M. Levitt, J. Mol. Biol. {\bf 249}, 493 (1995)
\bibitem{yyy} M. E. J. Newman, SIAM Rev. {\bf 45}, 167 (2003)
\bibitem{binder} K. Binder and A. P. Young, Rev. Mod. Phys. {\bf 58}, 801 (1986)
\bibitem{zzz} W. Li, Computers Chem. {\bf 21} 257 (1997)


%\bibitem{vanHove} L.\ van Hove,
%Physica \textbf{16}, 137 (1950) (reprinted in  \cite{lieb}, p.\ 28)
%\bibitem{lieb}
%E.\ H.\ Lieb and D.\ C.\ Mattis, eds., {\em Mathematical Physics
%in One Dimension} (Academic, New York, 1966).
%\bibitem{landau} L.\ D.\ Landau and E.\ M.\ Lifshitz, {\em Statistical
%Physics Part 1} (Pergamon, New York, 1980).
%\bibitem{us4} J.\ A.\ Cuesta and A.\ S\'anchez, J.\ Stat.\ Phys.\ {\bf 115}, 869 (2004).
%\bibitem{schiff} L.\ I.\ Schiff, {\em Quantum Mechanics} $2^{nd}$ edn.
%(McGraw-Hill, New York, 1968).
%\bibitem{us3} S.\ Ares, J.\ A.\ Cuesta, A.\ S\'anchez and R.\ Toral,
%Phys.\ Rev.\ E {\bf 67}, 046108 (2003).
%\bibitem{ch} R.\ Courant and D.\ Hilbert, {\em Methods of Mathematical Physics
%vol.\ I} (Wiley Classics, New York, 1989).
%\bibitem{theo} N.\ Theodorakopoulos, Phys.\ Rev.\ E {\bf 68}, 026109 (2003).


%\bibitem{Batrouni} G.\ G.\ Batrouni and T.\ Hwa, Phys.\ Rev.\ Lett. {\bf
%72}, 4133 (1994).
%\bibitem{else} S.\ Ares, A.\ S\'anchez and A.\ R.\ Bishop, unpublished.
%\bibitem{unpub} A.\ S\'anchez and A.\ R.\ Bishop, unpublished.

%\bibitem{Ising} M. Ya Azbel, Phys. Rev. A, {\bf 20}, 1671 (1979) and references therein.
%\bibitem{NN}J. Santa Lucia, Jr., Proc. Natl. Acad. Sci U.S.A. {\bf 95}, 1460 (1998).
%\bibitem{zipper} C. Kittel, Am. J. Phys. {\bf 37}, 917 (1969).
%\bibitem{Ivanov} V. Ivanov, Y. Zeng, and G. Zocchi, Phys. Rev. E {\bf 70}, 051907 (2004).
\end{thebibliography}
\end{document}